\documentclass[3p,times, 11pt]{elsarticle}
 \usepackage[latin1]{inputenc}
 \usepackage{hyperref}
\hypersetup{
    colorlinks=true,
    linkcolor=blue,
    citecolor=blue,
    urlcolor=blue
}
\usepackage{amssymb,amsmath,amsthm,amscd,latexsym}
\usepackage{mathrsfs}
\usepackage{mathrsfs}
\usepackage{amsfonts}
\usepackage{amsmath}
\usepackage{amssymb}
\usepackage{amscd}
\usepackage{xypic}
\usepackage{graphicx}
\usepackage{epstopdf}
\usepackage{yfonts}
\usepackage[T1]{fontenc}
\usepackage[all]{xy}
\usepackage{times}
\usepackage{color}
\usepackage{caption}

\usepackage[linesnumbered,ruled,vlined]{algorithm2e}
\usepackage[new]{old-arrows}
\newtheorem{Theorem}{Theorem}

\newtheorem{Definition and Notation}{Definition and Notation}
\newtheorem{Lemma}{Lemma}

\newtheorem{Proposition}{Proposition}
\newtheorem{Corollary}{Corollary}

{\theoremstyle{definition}

\newtheorem{Definition}{Definition}
\newtheorem{Example}{Example}[section]
\newtheorem{Remark}{Remark}

\journal{Discrete Mathematics}

\usepackage{amssymb}

\usepackage[figuresright]{rotating}

\begin{document}

\begin{frontmatter}

\title{On LCP codes over  a mixed ring alphabet }
\author[MU,1]{Maryam Bajalan}\ead{maryam.bajalan@math.bas.bg}
\fntext[1]{Maryam Bajalan is supported by the Bulgarian Ministry of Education and Science, Scientific Programme "Enhancing the Research Capacity in Mathematical Sciences (PIKOM)", No. DO1-67/05.05.2022.
}
\author[JC]{Javier de la Cruz}\ead{jdelacruz@uninorte.edu.co}
\author[UBe]{Alexandre Fotue Tabue}\ead{alexfotue@gmail.com}
\author[UVa,2]{Edgar Mart\'inez-Moro
}\ead{edgar.martinez@uva.es}
\fntext[2]{This author was supported in part by Grant TED2021-130358B-I00 funded by MCIN/AEI/10.13039/501100011033 and by the ?European Union NextGenerationEU/PRTR? }

\address[MU]{Institute of Mathematics and Informatics, Bulgarian Academy of Sciences, Bl. 8, Acad. G. Bonchev Str., 1113, Sofia, Bulgaria}
\address[JC]{Department of Mathematics, Universidad del Norte, Barranquilla, Colombia}
\address[UBe]{Department of Mathematics, University of Bertoua,  Cameroon}
\address[UVa]{Institute of Mathematics, University of Valladolid, Castilla, Spain}

\begin{abstract}
In this paper, we introduce a standard generator matrix for mixed-alphabet linear codes over finite chain rings. Furthermore, we show that, when one has a linear complementary pair (LCP) of mixed-alphabet linear codes, both codes are weakly-free.
Additionally, we establish that any mixed-alphabet product group code is separable. Thus, if one has a pair  $\{C, D\}$ of mixed-alphabet product group codes over a finite chain ring that forms a LCP, it follows that $C$ and the Euclidean dual of $D$ are permutation equivalent.
\end{abstract}


\end{frontmatter}

\section{Introduction}
Let $p$ be a prime and $r, s, \alpha, \beta$  be positive integers such
that $1\leq r\leq s$, a
 subgroups of the additive group
$(\mathbb{Z}_{p^r})^\alpha\times(\mathbb{Z}_{p^s})^\beta$ is
called a  $\mathbb{Z}_{p^r}\mathbb{Z}_{p^s}$-\textit{additive code}.  The authors of \cite{BFRR10} studied the
algebraic structure of $\mathbb{Z}_{2}\mathbb{Z}_{4}$-additive
codes and determined their generator matrices and duality. Those results were generalized to $\mathbb{Z}_{p^r}\mathbb{Z}_{p^s}$-additive codes in
\cite{AS15}. One can state this class of codes in a more general setting of mixed-alphabet  over a finite chain ring
$R$ as follows,  an $R\overline{R}$-linear code of block-length $(\alpha, \beta)$ is a submodule of the $R$-module
$R^\alpha\times\overline{R}^{\,\beta}$, where $\overline{R}$ is a
quotient ring of $R$ modulo an ideal of $R$.

A
pair of  linear codes $ {\{C, D\}}$ of
length $n$ over a finite field $\mathbb F$ is called  a \textit{linear complementary
pair} (LCP) of codes if $C\oplus D=\mathbb F^n$ and  $C\cap D=\{\mathbf 0\} $. In \cite{CG16, CGOOS18}, LCPs of codes over
finite fields have been studied due to their rich algebraic
structure{s} and wide applications in cryptography.
{On one hand} they showed that the LCPs can help
{with improvement} the security of the information
processed by sensitive devices, especially against side-channel
attacks (SCA) and fault injection attacks (FIA).
On the other hand, they established  {the security
parameter of a given LCPs pair $\{C, D\}$ over a finite field is}
$\min\{ d(C), d(D^\perp)\}$, where $d(C)$ denotes the minimum
Hamming distance for $C$.  {LCPs of
codes over finite rings in} have been studied in\cite{LL23a}. In this later paper, the authors gave a necessary
and sufficient condition for a  {pair} of linear
codes over finite rings to be LCP, and constructed a
maximum-distance-separable LCP of codes over {the}
ring $\mathbb{Z}_4$.

Borello et al. showed in \cite{BCW20} that for ${\{C, D\}}$ a LCP of linear
codes (ideals) in the group ring  $\mathbb{F}[G]$ (where $G$ is a finite group) the codes $C$ and $D^\perp$ are permutation
equivalent. In \cite{GMS20} it was extended Borello's
results to codes over finite chain
rings and to linear codes over finite Frobenius rings in \cite{LL23b}.

A
pair of mixed-alphabet linear codes $ {\{C, D\}}$ of
length $n$ over a finite chain ring $R$ is called  a \textit{linear complementary
pair} (LCP) of codes if $C\oplus D=\overline{R}^{\,\alpha}\times
R^\beta$ and they have a trivial intersection. In the case $D=C^\perp$ where $C^\perp$ is the Euclidean
dual of $C$, the code $C$ is called
\textit{linear complementary dual} (LCD), that is  $ {\{C,
C^\perp\}}$ is an LCP of codes. In \cite{BBDF20} LCD
$\mathbb{Z}_2\mathbb{Z}_4$-additive codes were studied. The results in
\cite{BFKM23} extended those results to finite chain
rings.

 {The main contribution of this paper  is characterizing
LCP mixed-alphabet codes over finite chain rings and, using that result,  extending
the results in \cite{GMS20} to mixed-alphabet linear codes. The outline of the paper is as follows}. Section~\ref{sec:1} provides an introduction to mixed-alphabet linear codes over chain rings as well as shows their standard generator matrix and a matrix for their duals based on it. In Section~\ref{Secc:2}, the problem of characterizing LCP codes over the direct product of finite chain rings is studied. Finally, in Subsection \ref{subsec31} the main result in \cite{BCW20} is extended to our alphabet.

\section{Mixed-alphabet linear codes}\label{sec:1}

Throughout this paper, $R$ denotes a finite commutative chain ring
with the maximal ideal $\texttt{J}(R)$ generated by $\theta$ with
the nilpotency index $s$ and $R^\times$  denote the multiplicative group
of $R$. Let  $q$ (a power of a prime number)  be the cardinality of the residue field of $R$.
It is well-known that there exists a unique subgroup
$\Gamma(R)\backslash\{0_R\}$ of $R^\times$ such that
$\Gamma(R)\backslash\{0_R\}\simeq(\mathbb{F}_q\backslash\{0_R\})$
and
$R^\times\simeq\Gamma(R)\backslash\{0_R\}\times(1+\texttt{J}(R))$
as groups. The set  $\Gamma(R)$ is called the
\textit{Teichm\"uller set} of $R$.   Also,  there exists a unique set of maps
$\left\{\gamma_t: R\rightarrow \Gamma(R)\;:\; 0\leq t<s\right\}$
such that
$x=\gamma_0(x)+\gamma_1(x)\theta+\gamma_2(x)\theta^2+\cdots+\gamma_{s-1}(x)\theta^{s-1},$
for any element $x$ in $R$. If we fix an integer $r$ such that  $1\leq r\leq s$, and we set
$\overline{R}:=R/\theta^rR$,  then the quotient ring $\overline{R}$
is the finite commutative chain ring  with a residue field
$\mathbb{F}_q$ and nilpotency index $r$, and the map
\begin{align}\label{pi}
\begin{array}{rlll}
  \pi : & R  & \longrightarrow & \overline{R} \\
          & x & \longmapsto & \overline{x}:=x+\theta^rR,
\end{array}
\end{align}
is a ring epimorphism. Thus, the map $\eta:
\Gamma(R)\backslash\{0_R\}\rightarrow\Gamma(\overline{R})\backslash\{0_{\overline{R}}\}$
defined as $\eta(x):=\overline{x}$ is a group isomorphism. Since
$\pi(\Gamma(R))=\Gamma(\overline{R})$ and by setting
$\eta(0_R)=0_{\overline{R}}$, the map $\eta$ can be extended to
$\Gamma(R)$ and  the group isomorphism $\eta$ defines the
injective map
\begin{align}\label{iota}
\begin{array}{clcl}
  \iota : & \overline{R} & \longhookrightarrow & R  \\
          & x & \longmapsto &
          \sum\limits_{t=0}^{r-1}\eta^{-1}(\gamma_t(x))\theta^t.
\end{array}
\end{align}
It can be easily checked that
$\pi\circ\iota=\texttt{Id}_{\overline{R}}$ and
$\theta^{r-s}\iota\circ\pi=\theta^{r-s}\texttt{Id}_R$.

Let $\alpha$ and $\beta$ be two positive integers. Let
$\overrightarrow{\textbf{0}}$  be the zero vector in
$R^{\alpha}\times\overline{R}^{\,\beta}$ and
$\overrightarrow{\textbf{u}}=(\textbf{u}\,|\,\overline{\textbf{u}})$
be an element of $R^{\alpha}\times\overline{R}^{\,\beta}$, where
$\textbf{u}\in R^\alpha$ and $\overline{\textbf{u}}\in
\overline{R}^{\,\beta}$. The additive group
$R^{\alpha}\times\overline{R}^{\,\beta}$ has an $R$-module
structure with the     multiplication $\ast
:R\times\left(R^{\alpha}\times\overline{R}^{\,\beta}\right)\longrightarrow
R^{\alpha}\times\overline{R}^{\,\beta}$ given by
\begin{equation}\label{opera}
    a\ast (u_1,\ldots, u_{\alpha}\mid \overline{u}_{\alpha+1},\ldots, \overline{u}_{\alpha+\beta})= \left(au_1,\ldots, au_{\alpha}\mid \overline{a}\,\overline{u}_{\alpha+1},\ldots, \overline{a}\,\overline{u}_{\alpha+\beta}\right).
\end{equation}

\begin{Definition} An \textit{$R\overline{R}$-linear code of block-length} $(\alpha,
\beta)$ is an $R$-submodule of $(R^\alpha\times\overline{R}^{\,\beta}, +, *)$.
\end{Definition}

In order to  establish a correspondence
between the set of $R\overline{R}$-linear codes of block-length
$(\alpha, \beta)$ and the set of $R$-linear codes of length
$\alpha+\beta$ contained in
$R^\alpha\times\overline{R}^{\,\beta}$, we introduce two maps
$\psi$ and $\chi$ defined as follows
\begin{equation}\label{eq:chi}
\begin{array}{ccc}
 \begin{array}{cccl}
 \chi : & \overline{R} & \longrightarrow & \theta^{s-r}R  \\
    & u & \longmapsto & \theta^{s-r}\iota(u)
\end{array}  & \text{ and } & \begin{array}{cccl}
  \psi : & \theta^{s-r}R &\longrightarrow & \overline{R}  \\
    & \theta^{s-r}u & \longmapsto & \pi(u).
\end{array}
\end{array}\end{equation} It is easy to see that, for any
$(u,v)$ in $R^2$, we have
$\chi(\pi(uv))=\chi(\overline{u})\iota(\pi(v))$.

\begin{Lemma}\label{lem-chi}
 The maps $\psi$ and $\chi$ defined in Equation~(\ref{eq:chi}) are $R$-module isomorphisms.  {Moreover},
$\psi^{-1}=\chi$.
\end{Lemma}

\begin{proof} Since $\pi\circ\iota=\texttt{Id}_{\overline{R}}$ and
$\theta^{r-s}\iota\circ\pi=\theta^{r-s}\texttt{Id}_R$, then  the maps
$\psi$ and $\chi$ are bijective and $\psi^{-1}=\chi$. Moreover,
for all $x$ and $y$ in $R$, we have
$$\theta^{s-r}\iota\circ\pi(x)+\theta^{s-r}\iota\circ\pi(y)=\theta^{s-r}\iota\circ\pi(x+y)$$
and $$y\theta^{s-r}\iota\circ\pi(x)=y\theta^{s-r}x=\theta^{s-r}
yx=\theta^{s-r}\iota\circ\pi(yx).$$ Thus, $\chi$ is an $R$-module
isomorphism. Since $\psi^{-1}=\chi$,  then it follows that $\psi$ is
also an $R$-module isomorphism.
\end{proof}

These maps $\chi$ and $\psi$ are naturally extended as follows
\begin{align}\label{chi1}
\begin{array}{lcll}
 \chi : & R^{\alpha}\times\overline{R}^{\,\beta} & \longrightarrow & R^{\,\alpha}\times \theta^{s-r}R^{\beta}  \\
    & \left(u_1,\ldots, u_{\alpha}\,|\,\overline{u}_{\alpha+1},\ldots, \overline{u}_{\alpha+\beta}\right) & \longmapsto & \left((u_1,\ldots, u_{\alpha}),\, (\theta^{s-r}\iota(\overline{u}_{\alpha+1}),\ldots,
    \theta^{s-r}\iota(\overline{u}_{\alpha+\beta}))\right),
\end{array}
\end{align} and
\begin{align}\label{psi1}
\begin{array}{lcll}
  \psi : & R^{\alpha}\times\theta^{s-r}R^{\beta} &\longrightarrow & R^{\alpha}\times \overline{R}^{\beta}  \\
    & \left((u_1,\ldots, u_{\alpha}),\,(\theta^{s-r} u_{\alpha+1},\ldots, \theta^{s-r} u_{\alpha+\beta})\right) & \longmapsto & \left(u_1,\ldots, u_{\alpha}\,|\,\pi(u_{\alpha+1}),\ldots,
    \pi(u_{\alpha+\beta})\right).
\end{array}
\end{align}

Note that Lemma~\ref{lem-chi} shows  that $C$ is an
$R\overline{R}$-linear code of block-length $(\alpha, \beta)$ if
and only if $\chi(C)$ is a linear code over $R$ of length
$\alpha+\beta$. Thus, the algebraic structure of any
$R\overline{R}$-linear code $C$ relies on the structure of $\chi(C)$.
The algebraic structure of linear codes over finite chain rings
has been studied in \cite{DS09, NS00}.

Let
$\overrightarrow{\textbf{v}}_1, \overrightarrow{\textbf{v}}_2,
\ldots, \overrightarrow{\textbf{v}}_k$ be nonzero codewords in $C$  an
$\overline{R}R$-linear code  of block-length $(\alpha, \beta)$.
The vectors $\chi(\overrightarrow{\textbf{v}}_1),
\chi(\overrightarrow{\textbf{v}}_2),\ldots,
\chi(\overrightarrow{\textbf{v}}_k)$ form an \emph{$R$-basis} for
$\chi(C)$, if $\chi(C)$ is the set of all $R$-linear combinations
of vectors $\chi(\overrightarrow{\textbf{v}}_1),
\chi(\overrightarrow{\textbf{v}}_2),\ldots,
\chi(\overrightarrow{\textbf{v}}_2),\ldots,
\chi(\overrightarrow{\textbf{v}}_k)$ and for any
$(\alpha_1,\ldots, \alpha_k)$ in $R^k$ such that
$\sum\limits_{i=1}^k\alpha_i\chi(\overrightarrow{\textbf{v}}_i)=\textbf{0}$
implies $\alpha_i\in\texttt{J}(R)$, for any $1\leq i\leq k$. Since
$\chi$ is an $R$-module isomorphism,
$\overrightarrow{\textbf{v}}_1, \overrightarrow{\textbf{v}}_2,
\ldots, \overrightarrow{\textbf{v}}_k$  form an $R$-basis for $C$,
if and only if $\chi(\overrightarrow{\textbf{v}}_1),
\chi(\overrightarrow{\textbf{v}}_2),\ldots,
\chi(\overrightarrow{\textbf{v}}_k)$ form an $R$-basis for
$\chi(C)$. Denote by $\textbf{M}_{k\times\alpha}(\overline{R})$
and $\textbf{M}_{k\times \beta}(R)$ the additive groups of
$(k\times\alpha)$-matrices over $\overline{R}$ and
$(k\times\beta)$-matrices over $R,$ respectively. A mixed-matrix
$\mathrm{G}$ in
$\textbf{M}_{k\times\alpha}(R)\times\textbf{M}_{k\times\beta}(\overline{R})$
is a \textit{generator matrix} for $C,$ if the rows of $\mathrm{G}$
form an $R$-basis for $C.$ From \cite[Proposition 3.2]{NS00}, a
generator matrix for $\chi(C)$ in  standard form is

\begin{align}\label{sf}\scriptsize{\left(
\begin{array}{ccccc|ccccc}
\mathrm{I}_{k_{0}} & \mathrm{G}_{0,1} & \mathrm{G}_{0,2} & \cdots & \mathrm{G}_{0,s-r} & \mathrm{O} & \theta^{s-r}\mathrm{G}_{0,s-r+1} & \theta^{s-r}\mathrm{G}_{0,s-r+2} & \cdots  & \theta^{s-r}\mathrm{G}_{0,s} \\
\mathrm{O} & \theta \mathrm{I}_{k_{1}} & \theta \mathrm{G}_{1,2} & \cdots & \theta \mathrm{G}_{1,s-r} & \mathrm{O} & \mathrm{O} & \theta^{s-r+1}\mathrm{G}_{1,s-r+2} & \cdots  & \theta ^{s-r+1}\mathrm{G}_{1,s} \\
\vdots  & \rotatebox{10}{$\ddots$}  & \rotatebox{10}{$\ddots$}  & \rotatebox{10}{$\ddots$}  & \vdots  & \vdots  & \vdots & \rotatebox{10}{$\ddots$} & \rotatebox{10}{$\ddots$}  & \vdots  \\
\mathrm{O} & \cdots  & \mathrm{O} & \theta^{s-r-1}\mathrm{I}_{k_{s-r-1}} & \theta^{s-r-1}\mathrm{G}_{s-r-1,s-r} & \mathrm{O} & \mathrm{O} & \cdots & \mathrm{O} & \theta ^{s-1}\mathrm{G}_{s-r-1,s} \\
\hline \mathrm{O} & \cdots  & \mathrm{O} & \mathrm{O} & \theta^{s-r}\mathrm{G}_{s-r,s-r} & \theta^{s-r}\mathrm{I}_{k_{s-r}} & \theta^{s-r}\mathrm{G}_{s-r,s-r+1} & \theta^{s-r}\mathrm{G}_{s-r,s-r+2} & \cdots  & \theta^{s-r}\mathrm{G}_{s-r,s} \\
\mathrm{O} & \cdots  & \mathrm{O} & \mathrm{O} & \theta ^{s-r+1}\mathrm{G}_{s-r+1,s-r} & \mathrm{O} & \theta ^{s-r+1}\mathrm{I}_{k_{s-r+1}} & \theta^{s-r+1}\mathrm{G}_{s-r+1,s-r+2} & \cdots  & \theta^{s-r+1}\mathrm{G}_{s-r+1,s} \\
\vdots  &  & \vdots  & \vdots  & \vdots  & \vdots  & \rotatebox{10}{$\ddots$}  & \rotatebox{10}{$\ddots$}  & \rotatebox{10}{$\ddots$}  & \vdots  \\
\mathrm{O} & \cdots  & \mathrm{O} & \mathrm{O} & \theta^{s-1}\mathrm{G}_{s-1,s-r} & \mathrm{O} & \cdots  &\mathrm{O} & \theta^{s-1}\mathrm{I}_{k_{s-1}} & \theta ^{s-1}\mathrm{G}_{s-1,s}%
\end{array}%
\right)\mathrm{U}},\end{align} where $\mathrm{O}$ is the all zero
matrix, $\mathrm{U}$ is a suitable permutation matrix, and the
columns are respectively grouped into blocks of sizes $k_0, k_1,
\ldots, k_{s-r-1}, \alpha-\mu, k_{s-r},\ldots, k_{s-1},$ and
$\beta-\rho$, where $\mu:=k_0+k_1+\cdots+ k_{s-r-1}$ and
$\rho:=k_{s-r}+k_{s-r+1}+\cdots+ k_{s-1}$.  Also $\mathrm{I}_{k_i}$
denotes  the identity matrix of size $k_i$, where $0\leq i< s-1$, and
\begin{itemize}
    \item for $0\leq i< j<s-r\leq h < l <s$, the size of the matrices $\mathrm{G}_{i,j}, \mathrm{G}_{h,l}$ and $\mathrm{G}_{i,h}$ is $k_i\times
k_j,$ $k_h\times k_l$ and $k_i\times k_l$, respectively;
    \item for $0\leq i <s-r\leq h <s$, the size of  the matrices $\mathrm{G}_{i,s-r}$ and $\mathrm{G}_{h,s-r}$ is $k_i\times
(\alpha-\mu),$ and $k_h\times (\alpha-\mu)$, respectively;
    \item for $0\leq i <s-r\leq h <s$, the size of the matrices $\mathrm{G}_{i,s}$ and $\mathrm{G}_{h,s}$ is $k_i\times
(\beta-\rho),$ and $k_h\times (\beta-\rho)$, respectively.
\end{itemize}
Of course, for any $1 \leq i<s,$ if $k_i = 0,$ then the
matrices $\mathrm{G}_{i,j}$ are suppressed in (\ref{sf}).
Since $\chi$ defined in (\ref{chi1}) is an $R$-module isomorphism,
it follows that $\mathrm{G}$ is a generator matrix for $C$ if and
only if $\chi(\mathrm{G})$ is a generator matrix for $\chi(C)$.
Therefore, we have the following.

\begin{Proposition}\label{propoG} Let $C$ be an $R\overline{R}$-linear code
whose $\chi(C)$ has a generator matrix as in Matrix (\ref{sf}).
Then $C$ is permutation equivalent to an $R\overline{R}$-linear
code with a generator matrix of the form
\begin{align}\label{sf0}
\footnotesize{\left(
\begin{array}{ccccc|ccccc}
\mathrm{I}_{k_{0}} & \mathrm{G}_{0,1} & \mathrm{G}_{0,2} & \cdots & \mathrm{G}_{0,s-r} & \mathrm{O} & \overline{\mathrm{G}}_{0,s-r+1} & \overline{\mathrm{G}}_{0,s-r+2} & \cdots  & \overline{\mathrm{G}}_{0,s} \\
\mathrm{O} & \theta \mathrm{I}_{k_{1}} & \theta \mathrm{G}_{1,2} & \cdots & \theta \mathrm{G}_{1,s-r} & \mathrm{O} & \mathrm{O} & \overline{\theta\mathrm{G}}_{1,s-r+2} & \cdots  & \overline{\theta\mathrm{G}}_{1,s} \\
\vdots  & \rotatebox{10}{$\ddots$}  & \rotatebox{10}{$\ddots$}  & \rotatebox{10}{$\ddots$}  & \vdots  & \vdots  & \vdots & \rotatebox{10}{$\ddots$} & \rotatebox{10}{$\ddots$}  & \vdots  \\
\mathrm{O} & \cdots  & \mathrm{O} & \theta^{s-r-1}\mathrm{I}_{k_{s-r-1}} & \theta^{s-r-1}\mathrm{G}_{s-r-1,s-r} & \mathrm{O} & \mathrm{O} & \cdots  & \mathrm{O} & \overline{\theta^{\,r-1}\mathrm{G}}_{s-r-1,s} \\
\hline \mathrm{O} & \cdots  & \mathrm{O} & \mathrm{O} & \theta^{s-r}\mathrm{G}_{s-r,s-r} & \mathrm{I}_{k_{s-r}} & \overline{\mathrm{G}}_{s-r,s-r+1} & \overline{\mathrm{G}}_{s-r,s-r+2} & \cdots  & \overline{\mathrm{G}}_{s-r,s} \\
\mathrm{O} & \cdots  & \mathrm{O} & \mathrm{O} & \theta^{s-r+1}\mathrm{G}_{s-r+1,s-r} & \mathrm{O} & \overline{\theta} \mathrm{I}_{k_{s-r+1}} & \overline{\theta\mathrm{G}}_{s-r+1,s-r+2} & \cdots  & \overline{\theta\mathrm{G}}_{s-r+1,s} \\
\vdots  &  & \vdots  & \vdots  & \vdots  & \vdots  & \rotatebox{10}{$\ddots$} & \rotatebox{10}{$\ddots$}  & \rotatebox{10}{$\ddots$}  & \vdots  \\
\mathrm{O} & \cdots  & \mathrm{O} & \mathrm{O} & \theta^{s-1}\mathrm{G}_{s-1,s-r} & \mathrm{O} & \cdots  & \mathrm{O} & \overline{\theta}^{\,r-1}\mathrm{I}_{k_{s-1}} & \overline{\theta^{\,r-1}\mathrm{G}}_{s-1,s}%
\end{array}%
\right)\mathrm{U},}\end{align} where $\mathrm{U}$ is a suitable
permutation matrix. Moreover, there is a unique $(s + 2)$-tuple of
nonnegative integers $(\alpha,\beta; k_0,\ldots, k_{s-r-1}\,|\,
k_{s-r},\ldots,k_{s-1})$, so-called the \emph{type} of $C,$ such
that $C$ isomorphic to the $R$-module
$\prod\limits_{\substack{t=0 \\
k_t\neq 0}}^{s-1}\left(R/\langle
\theta^{\,s-t}\rangle\right)^{k_t}.$ Moreover,
$\log_q(|C|)=\sum\limits_{t=0}^{s-1}(s-t)k_t$, so-called dimension
of the code $C$, and denoted by $\dim(C)$.
\end{Proposition}

\begin{Definition} An $R\overline{R}$-linear code $C$ is \textit{separable}, if $C = C_1
\times C_1$, where $C_1$ is linear over $R$ and $\overline{C}_2$
is linear over $\overline{R}$. Note that a generator matrix for an
$R\overline{R}$-linear separable code $C_1 \times C_1$ is
$$\left(
\begin{array}{c|c}
  \mathrm{G}_1 & \mathrm{O} \\
  \mathrm{O} & \overline{\mathrm{G}}_2
\end{array}
\right),$$ where $\mathrm{G}_1$ is a generator matrix for $C_1$ and
$\overline{\mathrm{G}}_2$ is a generator matrix for $\overline{C}_2$.\end{Definition}

\begin{Example} Let $R^{\alpha}\times\overline{R}^{\,\beta}$ be the ambient space. Then $R^{\alpha}\times\overline{R}^{\,\beta}$ is an
$R\overline{R}$-linear code with a standard generator matrix
$$\left(
\begin{array}{c|c}
 \mathrm{I}_\alpha & \mathrm{O}  \\
  \mathrm{O} & \mathrm{I}_\beta
\end{array}%
\right).$$ Thus its type is $(\alpha,\beta; \alpha, 0, \ldots,
0\,|\,\beta, 0,\ldots, 0).$ Obviously, if $\beta\neq 0$, then
$R^{\alpha}\times\overline{R}^{\,\beta}$ is not free as an
$R$-module, since
$\chi(R^{\alpha}\times\overline{R}^{\,\beta})=R^{\alpha}\times
\theta^{s-r}R^{\,\beta}$ and $\chi$ is  {an} $R$-module isomorphism.
\end{Example}

This fact motivates the following definition.

\begin{Definition} An $R\overline{R}$-linear code of type $(\alpha,\beta; k_0,\ldots,
k_{s-r-1}\,|\, k_{s-r},\ldots,k_{s-1})$ is \textit{weakly-free}, if
$k_1=\cdots=k_{s-r-1}=k_{s-r+1}=\cdots=k_{s-1}=0.$
\end{Definition}

\begin{Example}\label{ex-gen} Let $C$ be a $\mathbb{Z}_8\mathbb{Z}_4$-linear code with a generator matrix
$ \left(
\begin{array}{cccc|ccc}
    7 & 6 & 5 & 4 & 1 & 2 & 3 \\
    6 & 4 & 0 & 2 & 2 & 0 & 1 \\
    4 & 4 & 2 & 4 & 0 & 1 & 2 \\
    2 & 6 & 6 & 2 & 1 & 0 & 1
\end{array}
\right). $ Then a generator matrix for $\chi(C)$ is $ \left(
\begin{array}{ccccccc}
    7 & 6 & 5 & 4 & 2 & 4 & 6 \\
    6 & 4 & 0 & 2 & 4 & 0 & 2 \\
    4 & 4 & 2 & 4 & 0 & 2 & 4 \\
    2 & 6 & 6 & 2 & 2 & 0 & 2
\end{array}
\right). $ After applying necessary row   {operations, }  we obtain the
following standard generator matrix of $\chi(C)$: $ \left(
\begin{array}{ccccccc}
    1 & 0 & 3 & 2 & 0 & 0 & 0 \\
    0 & 6 & 6 & 0 & 2 & 0 & 0 \\
    0 & 4 & 2 & 0 & 0 & 2 & 0 \\
    0 & 0 & 2 & 0 & 0 & 0 & 2
\end{array}
\right). $ Thus a standard generator matrix of $C$ is $ \left(
\begin{array}{cccc|ccc}
    1 & 0 & 3 & 2 & 0 & 0 & 0 \\
    0 & 6 & 6 & 0 & 1 & 0 & 0 \\
    0 & 4 & 2 & 0 & 0 & 1 & 0 \\
    0 & 0 & 2 & 0 & 0 & 0 & 1
\end{array}
\right)$ and its type  is $(4, 3; 1\,|\, 3, 0)$. We can see that
$C$ is weakly-free.
\end{Example}

\begin{Remark} The standard generator matrix for any weakly-free $R\overline{R}$-linear
code $C$ of type $$(\alpha,\beta; \mu, 0, \ldots, 0\,|\,\rho,
0,\ldots, 0)$$ is of the form
   \begin{align}\label{wf-mat}\left(
\begin{array}{cc|cc}
\mathrm{I}_{\mu} & \mathrm{G}_{1,1} & \mathrm{O} & \overline{\mathrm{G}}_{1,2} \\
\hline \mathrm{O} & \theta^{s-r}\mathrm{G}_{2,1} &
\mathrm{I}_{\rho} & \overline{\mathrm{G}}_{2,2}
\end{array}
\right)\mathrm{U},
\end{align}
where $\mathrm{U}$ is a suitable permutation matrix. Note that $C$
is a free code if and only if $\rho=0$.
\end{Remark}

\begin{Lemma}\label{weakly} Let $\{C,\, \,D\}$ be a pair of $R\overline{R}$-linear codes of block length $(\alpha, \beta)$
such that $C\cap D=\{\overrightarrow{\textbf{0}}\}$. If $C\oplus
D$ is weakly-free code, then both $C$ and $D$ are also weakly-free codes.
\end{Lemma}

\begin{proof} Let $(\alpha,\beta; k_0,\ldots, k_{s-r-1}\,|\,
k_{s-r},\ldots,k_{s-1})$ and $(\alpha,\beta; k'_0,\ldots,
k'_{s-r-1}\,|\, k'_{s-r},\ldots,k'_{s-1})$ be the types of $C$
and $D$ respectively. Now $C\cap
D=\{\overrightarrow{\textbf{0}}\}$ and $C\oplus D$ is weakly,
therefore the type of $C\cap D$ is $$(\alpha,\beta; \mu,0,\ldots,
0\,|\, \rho, 0,\ldots,0).$$   {Since the codes $C$ and $D$ are submodules}
of $C\oplus D$, it follows that
 $k_t=k'_t\leq 0$ for any $1\leq t \leq s-r-1$ and
$k_{s-r+t}=k'_{s-r+t}\leq 0$ for any $1\leq t \leq r-1$. Thus the
types of $C$ and $D$ are $(\alpha,\beta; k_0, 0,\ldots, 0\,|\,
k_{s-r}, 0,\ldots,0)$ and $(\alpha,\beta; k'_0, 0,\ldots, 0\,|\,
k'_{s-r}, 0,\ldots,0)$, respectively. Hence $C$ and $D$ are
weakly-free codes.
\end{proof}

 \subsection{Duality}
In order to define a duality of $R\overline{R}$-linear codes of block-length
$(\alpha, \beta)$, we consider the following bilinear map, usually called  the \textit{inner
product} on $R^{\alpha}\times\overline{R}^{\,\beta}$
$$\begin{array}{rlll}
 [ -\, ,\,- ] : & \left(R^{\alpha}\times\overline{R}^{\,\beta}\right)^2 & \longrightarrow & R \\
    & \left((\textbf{u}\,|\,\overline{\textbf{u}})\,,\,(\textbf{v}\,|\,\overline{\textbf{v}})\right) & \longmapsto & \langle \textbf{u},\textbf{v}\rangle_R+\chi(\langle \overline{\textbf{u}},
    \overline{\textbf{v}}\rangle_{\overline{R}}),
\end{array}$$
where $\langle u, v\rangle_R:=u_1v_1+\cdots+ u_{\alpha}v_{\alpha}$
and $\langle \overline{u},
\overline{v}\rangle_{\overline{R}}:=\overline{u}_{\alpha+1}
\overline{v}_{\alpha+1}\cdots+
\overline{u}_{\alpha+\beta}\overline{v}_{\alpha+\beta}$. For any
$R\overline{R}$-linear code $C$ of block-length $(\alpha, \beta)$,
the \emph{dual code} of $C,$ denoted by $C^{\perp},$ is defined as
\begin{equation}
C^{\perp}:=\left\{\overrightarrow{\textbf{u}}\in R^{\alpha}\times\overline{R}^{\,\beta}\mid [\overrightarrow{\textbf{u}}\,,\,\overrightarrow{\textbf{v}}]=0, \text{ for all } \overrightarrow{\textbf{v}}\in C\right\}.
\end{equation}
Note that $C^{\perp}$ is also an $R\overline{R}$-linear code.

The inner product on $R^{\alpha}\times\overline{R}^{\,\beta}$ can be expressed as the Euclidian inner
product on $R^{\alpha+\beta}$, for that we extend the map $\iota$ defined
in (\ref{iota}) to $R^{\alpha}\times \overline{R}^{\beta}$ in the
following way and denote the extension as $\iota$ once again:
\begin{align}\label{iota1}
\begin{array}{llll}
   \iota : & R^{\alpha}\times \overline{R}^{\beta} & \longhookrightarrow & R^{\alpha+\beta}  \\
    & (u_1,\ldots, u_{\alpha}\,|\,\overline{u}_{\alpha+1},\ldots, \overline{u}_{\alpha+\beta}) & \longmapsto & \left(u_1,\ldots, u_{\alpha}, \iota(\overline{u}_{\alpha+1}),\ldots,
    \iota(\overline{u}_{\alpha+\beta})\right).
\end{array}
\end{align}
So, we have this equality.

\begin{Lemma}\label{inner}
Let $(\overrightarrow{\textbf{u}},\overrightarrow{\textbf{v}})$ in
$\left(R^\alpha \times\overline{R}^{\,\beta}\right)^2$. Then
 $$\left[\overrightarrow{\textbf{u}},\overrightarrow{\textbf{v}}\right]=\left\langle
\iota(\overrightarrow{\textbf{u}})\,,\,\chi(\overrightarrow{\textbf{v}})\right\rangle_R=\left\langle
\chi(\overrightarrow{\textbf{u}})\,,\,\iota(\overrightarrow{\textbf{v}})\right\rangle_R,$$
where $\langle -\,,\,-\rangle_R$ is the standard inner product on
$R^{\alpha+\beta}.$
\end{Lemma}

\begin{proof} Set $\overrightarrow{\textbf{u}}=(u_1,\ldots, u_{\alpha}\,|\,\overline{u}_{\alpha+1},\ldots, \overline{u}_{\alpha+\beta})$, and $\overrightarrow{\textbf{v}}=(v_1,\ldots, v_{\alpha}\,|\,\overline{v}_{\alpha+1},\ldots,
\overline{v}_{\alpha+\beta})$. We have
\begin{eqnarray*}
  \left[ \overrightarrow{\textbf{u}},\overrightarrow{\textbf{v}}\right] &=& \sum\limits_{i=1}^{\alpha}u_iv_i+\chi\left(\sum\limits_{i=\alpha+1}^{\alpha+\beta}\overline{u}_i\overline{v}_i\right); \\
    &=& \sum\limits_{i=1}^{\alpha}u_iv_i+\sum\limits_{i=\alpha+1}^{\alpha+\beta}\chi\left(\overline{u}_i\overline{v}_i\right), \text{ since $\chi$ is an $R$-module isomorphism}; \\
    &=& \sum\limits_{i=1}^{\alpha}u_iv_i+\sum\limits_{i=\alpha+1}^{\alpha+\beta}\chi(\overline{u}_i)\iota(\overline{v}_i), \text{ since $\chi\left(\overline{u}_i\overline{v}_i\right)=\iota(\overline{u}_i)\chi(\overline{v}_i)$};\\
    &=& \left\langle \iota(\overrightarrow{\textbf{u}})\,,\,\chi(\overrightarrow{\textbf{v}})\right\rangle_R.
\end{eqnarray*}
Likewise, we have $\left[
\overrightarrow{\textbf{u}},\overrightarrow{\textbf{v}}\right]=\left\langle
\chi(\overrightarrow{\textbf{u}})\,,\,\iota(\overrightarrow{\textbf{v}})\right\rangle_R$.
\end{proof}
Let us define the map  $\varphi$   as follows:\begin{align}\label{iota1}
\begin{array}{cccc}
  \varphi : & R^{\alpha+\beta} & \longrightarrow & R^{\alpha}\times \overline{R}^{\,\beta}  \\
    & (u_1,\ldots, u_{\alpha}, u_{\alpha+1},\ldots, u_{\alpha+\beta}) & \longmapsto & \left(u_1,\ldots, u_{\alpha}\,|\,\pi(u_{\alpha+1}),\ldots,
    \pi(u_{\alpha+\beta})\right).
\end{array}
\end{align}
Note that $\varphi\circ\iota=\texttt{Id}_{R^{\alpha}\times
\overline{R}^{\,\beta}}$.

\begin{Lemma}\label{dualcode}
 If $C$ is an $R\overline{R}$-linear code of block-length
$(\alpha,\beta)$, then $C^\perp=\varphi(\chi(C)^\perp)$.
\end{Lemma}

\begin{proof}
\begin{eqnarray*}
  \varphi(\chi(C)^\perp) &=& \left\{\varphi(\textbf{u})\in R^{\alpha}\times \overline{R}^{\,\beta}\,:\,\textbf{u}\in\chi(C)^\perp\right\}; \\
    &=& \left\{\varphi(\textbf{u})\in R^{\alpha}\times \overline{R}^{\,\beta}\,:\,\left(\forall \overrightarrow{\textbf{c}}\in C\right)\left(\left\langle\textbf{u}\,,\,\chi(\overrightarrow{\textbf{c}})\right\rangle_R=0\right)\right\}, \text{ by the definition of the dual code}; \\
    &=& \left\{\varphi(\iota(\overrightarrow{\textbf{w}})+\theta^r(\textbf{0}\,|\,\textbf{u}'))\in R^{\alpha}\times \overline{R}^{\,\beta}\,:\,\left(\forall \overrightarrow{\textbf{c}}\in C\right)\left(\left\langle\iota(\overrightarrow{\textbf{w}})+\theta^r(\textbf{0}\,|\,\textbf{u}')\,,\,\chi(\overrightarrow{\textbf{c}})\right\rangle_R=0\right)\right\},\\
    & & \text{ since $\textbf{u}=\iota(\overrightarrow{\textbf{w}})+\theta^r(\textbf{0}\,|\,\textbf{u}')$, where $\textbf{u}'\in R^\beta$}; \\
    &=& \left\{\varphi(\iota(\overrightarrow{\textbf{w}}))\in R^{\alpha}\times \overline{R}^{\,\beta}\,:\,\left(\forall \overrightarrow{\textbf{c}}\in C\right)\left(\left\langle\iota(\overrightarrow{\textbf{w}}))\,,\,\chi(\overrightarrow{\textbf{c}})\right\rangle_R=0\right)\right\},\\
    & & \text{ since $\varphi(\iota(\overrightarrow{\textbf{w}})+\theta^r(\textbf{0}\,|\,\textbf{u}'))=\varphi(\iota(\overrightarrow{\textbf{w}}))$ and $\left\langle\iota(\overrightarrow{\textbf{w}})+\theta^r(\textbf{0}\,|\,\textbf{u}')\,,\,\chi(\overrightarrow{\textbf{c}})\right\rangle_R=\left\langle\iota(\overrightarrow{\textbf{w}})\,,\,\chi(\overrightarrow{\textbf{c}})\right\rangle_R$}; \\
    &=& \left\{ \overrightarrow{\textbf{w}}\in R^{\alpha}\times \overline{R}^{\,\beta} \,:\,\left(\forall \overrightarrow{\textbf{c}}\in C\right)\left(\left\langle\iota(\overrightarrow{\textbf{w}})\,,\,\chi(\overrightarrow{\textbf{c}})\right\rangle_R=0\right)\right\}, \text{ since $\varphi\circ\iota=\texttt{Id}_{R^{\alpha}\times\overline{R}^{\,\beta}}$ }; \\
    &=& C^\perp.
\end{eqnarray*}
\end{proof}

Let $C$ be an $R\overline{R}$-linear weakly-free code of
block-length $(\alpha,\beta)$ with a generator matrix $\mathrm{G}$
in the standard form as in (\ref{wf-mat}). Then $\chi(C)$ is an
$R$-linear code of length $\alpha+\beta$ and by \cite[Theorem
3.10]{NS00}, the parity-check matrix for $\chi(C)$ is given by
\begin{align}
 \left(
\begin{array}{cccc}
  -\mathrm{G}_{1,1}^{\texttt{tr}} & \mathrm{I}_{\alpha-\mu} & -\mathrm{G}_{2,1}^{\texttt{tr}} &  \mathrm{O} \\
  -\theta^{s-r}\mathrm{G}_{1,2}^{\texttt{tr}} & \mathrm{O} & -\mathrm{G}_{2,2}^{\texttt{tr}} &  \mathrm{I}_{\beta-\rho} \\
  \mathrm{O} & \mathrm{O} & \theta^r\mathrm{I}_{\rho} & \mathrm{O}
\end{array}%
\right)(\mathrm{U}^{-1})^{\texttt{tr}},
\end{align} where $\texttt{tr}$ denotes the transpose matrix.
From Lemma \ref{dualcode}, and the facts that
$\chi(C^\perp)\subseteq \iota(C^\perp)$ and $\varphi\circ\iota =
\texttt{Id}_{R^\alpha\times\overline{R}^{\,\beta}},$ we have the following result.

\begin{Proposition}\label{gen-check}
 If  $C$ is an $R\overline{R}$-linear weakly-free code with
a generator matrix $\mathrm{G}$ in standard form as in
(\ref{wf-mat}), then a parity-check matrix for $C$ is given by
\begin{align}\label{gc0}
 \left(
\begin{array}{cc|cc}
  -\mathrm{G}_{1,1}^{\texttt{tr}} & \mathrm{I}_{\alpha-\mu} & -\overline{\mathrm{G}_{2,1}}^{\texttt{tr}} &  \mathrm{O} \\
  \hline
  -\theta^{s-r}\mathrm{G}_{1,2}^{\texttt{tr}} & \mathrm{O} & -\overline{\mathrm{G}_{2,2}}^{\texttt{tr}} &  \mathrm{I}_{\beta-\rho}
\end{array}%
\right)(\mathrm{U}^{-1})^{\texttt{tr}},
\end{align}
and the type of $C^\perp$ is
$\left(\alpha,\beta;\alpha-\mu,0,\ldots,0\,|\, \beta-\rho, 0,
\ldots,0\right).$ Moreover, $\chi(C^\perp)\subseteq\chi(C)^\perp$.
\end{Proposition}

\begin{Example}\label{ex-parity} Let $C$ be the $\mathbb{Z}_8\mathbb{Z}_4$-linear code with a standard generator matrix $ \left(
\begin{array}{cccc|ccc}
    1 & 0 & 3 & 2 & 0 & 0 & 0 \\
    0 & 6 & 6 & 0 & 1 & 0 & 0 \\
    0 & 4 & 2 & 0 & 0 & 1 & 0 \\
    0 & 0 & 2 & 0 & 0 & 0 & 1
\end{array}
\right)$. A parity-check matrix for $C$ is $\left(
\begin{array}{cccc|ccc}
    0 & 1 & 0 & 0 & 1 & 2 & 0 \\
    5 & 0 & 1 & 0 & 1 & 3 & 3 \\
    6 & 0 & 0 & 1 & 0 & 0 & 0
\end{array}
\right)$.
\end{Example}

\begin{Corollary}\label{sep-dual} If  $C$ is an $R\overline{R}$-additive code,
 then $C$ is separable if and only if $C^\perp$ is separable.
Moreover, if $C$ is separable and $C=C_1\times \overline{C}_2$,
then $C^\perp = C_1^\perp\times \overline{C}_2^\perp.$
\end{Corollary}

\begin{Corollary}\label{cor-dual}
If $C$ is an $R\overline{R}$-linear weakly-free code of
block-length $(\alpha, \beta)$, then
\begin{enumerate}
    \item $C^\perp$ is weakly-free;
    \item $\dim(C)+\dim(C^{\perp})=s\alpha+r\beta,$ and $(C^\perp)^\perp=C.$
\end{enumerate}
\end{Corollary}

\begin{Corollary}\label{sum-inter}
Let $C$ and $D$ be $R\overline{R}$-linear codes of block-length
$(\alpha, \beta)$. Then $(C+D)^{\perp}=C^{\perp}\cap D^{\perp}.$
\end{Corollary}

\section{LCP of codes over a direct product of finite chain rings}\label{Secc:2}

\subsection{Characterization of LCP codes}
From now on   $\{C,  D\}$ will denote a pair of $R\overline{R}$-linear
codes of block-length $(\alpha, \beta)$.

\begin{Definition} The pair $\{C,  D\}$ is a \textit{linear
complementary pair} (LCP) of codes, if
$C\cap D=\{\overrightarrow{\textbf{0}}\}$ and
$C+D=R^{\alpha}\times\overline{R}^{\,\beta}$.
\end{Definition}

 From Lemma~\ref{weakly} we have that  $C$ and $D$ are weakly-free codes, if $\{C, D\}$ is an
LCP of codes.

\begin{Proposition}\label{dual-lpc} The following statements  are
equivalent.
\begin{enumerate}
    \item $\{C, D\}$ is an LCP of codes.
    \item $\{C^{\perp},D^{\perp}\}$ is an LCP of codes.
\end{enumerate}
\end{Proposition}

\begin{proof} By definition of LCP, we have
$\{C,  D\} \,\, \text{ is LCP }$ means
$C+D=R^{\alpha}\times\overline{R}^{\beta}$ and $C\cap
D=\{\overrightarrow{\textbf{0}}\}$. Now,
$(R^{\alpha}\times\overline{R}^{\beta})^{\perp}=\{\overrightarrow{\textbf{0}}\}$
and
$\{\overrightarrow{\textbf{0}}\}^{\perp}=R^{\alpha}\times\overline{R}^{\beta}$.
From Corollary \ref{sum-inter}, $(C+D)^\perp=C^\perp\cap D^\perp$
and $C^\perp\cap D^\perp=C^\perp+D^\perp$. Thus $\{C, D\}$ is
LCP, if and only if $\{C^{\perp},D^{\perp}\}$ is LCP.
\end{proof}

\begin{Lemma}\label{lem-dim} Let $C,D$ be two  $R\overline{R}$-linear
codes, then  $$\dim\left(C+D\right)= \dim\left(C\right)+ \dim\left(D\right)- \dim\left(C\cap
D\right).$$
\end{Lemma}

\begin{proof}
The map $g:  C\times D   \longrightarrow   C+D$ defined by
$g(\overrightarrow{\textbf{u}},\overrightarrow{\textbf{v}})=\overrightarrow{\textbf{u}}+\overrightarrow{\textbf{v}}$
is an $R$-modules epimorphism. By the first isomorphism theorem,
it follows that $C\times D/\texttt{Ker}(g)\cong  C+D $ (as
$R$-modules). Now, $C\cap D$ and $\texttt{Ker}(g)$ are isomorphic,
since $\texttt{Ker}(g)=\{(\overrightarrow{\textbf{u}},
-\overrightarrow{\textbf{u}})\;:\; \overrightarrow{\textbf{u}}\in
C\cap D\}$. Therefore, $\left|C\times
D/\texttt{Ker}(g)\right|=\left|C\times D\right|/\left|C\cap
D\right|=|C+D|$. Thus $\dim\left(C+D\right)= \dim\left(C\right)+
\dim\left(D\right)- \dim\left(C\cap D\right).$
\end{proof}
From this lemma, note that   $\{C, \,D\}$ is LCP, if and only
if $\dim(C+D)=\dim(C)+\dim(D)=s\alpha+r\beta$.

\begin{Proposition}\label{proper2}
   If $\{C, D\}$ is a LCP of codes, then  $\dim(D^\perp)=\dim(C)$.
\end{Proposition}

\begin{proof}
Assume then $\{C, D\}$ is LCP. It follows that
$\dim(C)+\dim(D)=s\alpha+r\beta$. From Proposition
\ref{gen-check}, $\dim(D)+\dim(D^\perp)=s\alpha+r\beta$. Hence,
$\dim(D^\perp)=\dim(C)$.
\end{proof}

\begin{Lemma}\label{sep-lcp} Let $\{C_1,\,D_1\}$ be a pair of
linear codes over $R$, and $\{\overline{C}_2,\,\overline{D}_2\}$
be a pair of linear codes over $\overline{R}$.  Then $\{C_1\times
\overline{C}_2,\, D_1\times \overline{D}_2\}$ is LCP if and only
if $\{C_1,\, D_1\}$ and $\{\overline{C}_2,\, \overline{D}_2\}$ are
LCP.
\end{Lemma}

\begin{proof} Since $$(C_1\times
\overline{C}_2)\oplus(D_1\times \overline{D}_2)=(C_1\oplus
D_1)\times(\overline{C}_2\oplus\overline{D}_2),$$ it follows that
$R^\alpha\times \overline{R}^{\,\beta}=(C_1\times
\overline{C}_2)\oplus(D_1\times \overline{D}_2)$ is logically
equivalent to $C_1\oplus D_1=R^\alpha$ and
$\overline{C}_2\oplus\overline{D}_2=\overline{R}^{\,\beta}$.
\end{proof}

\begin{Remark}\cite[Coroallary 2.4.]{LL15}\label{inv}
Let $\mathrm{M}$ be an $(\alpha+\beta)\times (\alpha+\beta)$
matrix with entries in the chain ring $R$ and
$\overrightarrow{\textbf{v}}$ be a vector in
$R^{\alpha}\times\overline{R}^{\,\beta}.$ If $\mathrm{M}$ is
non-singular, then the linear system of equations
$\chi(\overrightarrow{\textbf{v}})\mathrm{M}^{\texttt{tr}}=\textbf{0}$
over $R$ has only the zero solution.
\end{Remark}

\begin{Theorem}\label{thm1}
Let $C$ and $D$ be two $R\overline{R}$-linear weakly-free codes
with generator matrices $\mathrm{G}_C$ and $\mathrm{G}_D,$
respectively. Define the matrix $\mathrm{G}$ to be
\begin{equation*}
  \mathrm{G}=  \left(\begin{array}{c}
  \mathrm{G}_C \\
  \mathrm{G}_D
\end{array}\right).
\end{equation*}
The pair $\{C, D\}$ is a LCP of codes if and only if the
following statements are simultaneously true:
\begin{enumerate}
    \item $\iota(\mathrm{G})$ is a non-singular matrix with entries in $R$.
    \item $\dim(C)+\dim(D)=s\alpha+r\beta.$
\end{enumerate}
\end{Theorem}

\begin{proof}
Suppose that $C\oplus D=R^{\alpha}\times \overline{R}^{\,\beta}$
is  a LCP of codes. Lemma~\ref{lem-dim} states
$\dim(C+D)=\dim(C)+\dim(D)=s\alpha+r\beta$, thus the matrix
$\mathrm{G}$ is a generator matrix for $C\oplus D$. Since $C\oplus
D=R^{\alpha}\times \overline{R}^{\,\beta}$, the matrix
$\iota(\mathrm{G})$ is a generator matrix for $R^{\alpha+\beta}$.
Now, the identity matrix $\mathrm{I}_{\alpha+\beta}$ is a
generator matrix for $R^{\alpha+\beta}$. Thus $\iota(\mathrm{G})$
is a square matrix with the rank $\alpha+\beta$, and there exists
a non-singular  matrix $\mathrm{P}$ over $R$ such that
$\mathrm{P}\iota(\mathrm{G})= \mathrm{I}_{\alpha+\beta}$. Hence,
$\iota(\mathrm{G})$ is a non-singular matrix over $R$.
{Moreover, Lemma \ref{lem-dim}  establishes the
validity of Statement~2.}

Conversely, if we take into account Corollary \ref{cor-dual}, we get
$\dim(C)=s\alpha+r\beta-\dim(C^{\perp})$ and
$\dim(D)=s\alpha+r\beta-\dim(D^{\perp}).$ 
Thus, $s\alpha+r\beta  =
\dim(C)+ \dim(D) =  \left(s\alpha+r\beta-\dim(C^{\perp})\right)+
\left(s\alpha+r\beta-\dim(D^{\perp})\right).$ Hence,
\begin{equation}\label{5}
    \dim(C^{\perp})+\dim(D^{\perp})=s\alpha+r\beta.
\end{equation}
We will prove  $C^{\perp}\cap
D^{\perp}=\{\overrightarrow{\textbf{0}}\}$. Suppose that
$\overrightarrow{\textbf{u}}\in C^{\perp}\cap D^{\perp}.$ So,
$[\overrightarrow{\textbf{u}}, \overrightarrow{\textbf{v}}_1]=0=[
\overrightarrow{\textbf{u}}, \overrightarrow{\textbf{v}}_2]$, for
any row $\overrightarrow{\textbf{v}}_1$ of $\mathrm{G}_C$ and any
row $\overrightarrow{\textbf{v}}_2$ of $\mathrm{G}_D.$ This gives
$[\overrightarrow{\textbf{u}}, \overrightarrow{\textbf{v}}]=0$,
for any row $\overrightarrow{\textbf{v}}$ of $\mathrm{G}.$  By
Lemma \ref{inner}, $\langle\chi(\overrightarrow{\textbf{u}}),
\iota\left(\overrightarrow{\textbf{v}}\right)\rangle_R=0.$ Note
that, we have proved
$\chi(\overrightarrow{\textbf{u}})\left(\iota\left(\mathrm{G}\right)\right)^{\texttt{tr}}=0.$
We know $\chi(\overrightarrow{\textbf{u}})$ is an element in
$R^{\alpha+\beta}$ and $\iota\left(\mathrm{G}\right)$ is a
non-singular matrix with entries in $R.$ By Remark \ref{inv}, it
follows that $\chi(\overrightarrow{\textbf{u}})=0$. Since $\chi$
is an $R$-module isomorphism, it follows that
$\overrightarrow{\textbf{u}}=\overrightarrow{\textbf{0}}$.

Lemma \ref{lem-dim} with Equation \eqref{5} and the above claim show
that $\{C^{\perp}, D^{\perp}\}$ is LCP. Now, the proof of
sufficiency is completed by Proposition \ref{dual-lpc}.
\end{proof}
\subsection{Mixed-alphabet product
group codes over finite chain ring}\label{subsec31}

Let $H$ and $K$ be two finite multiplicative groups of
order $\alpha$ and $\beta$, respectively. Denote the elements of the groups as $H := \{h_1,
h_2,\ldots, h_\alpha\}$ and $K := \{k_{\alpha+1},
k_{\alpha+2},\ldots, k_{\alpha+\beta}\}.$ The sets $R[H]$ and
$\overline{R}[K]$ given by :
$$R[H]:=\left\{\sum\limits_{i=1}^{\alpha}u_ih_i\;:\;(u_1, \ldots,
u_\alpha)\in R^\alpha\right\} \quad \text{ and
}\quad\overline{R}[H]:=\left\{\sum\limits_{i=\alpha+1}^{\alpha+\beta}u_ih_i\;:\;(u_{\alpha+1},
\ldots, u_{\alpha+\beta})\in \overline{R}^\beta\right\}.$$  have an algebraic structure of rings,
known as group rings. Moreover, $R[H]$ and $\overline{R}[K]$ are
free $R$-modules with basis $H$ and $K$, respectively. Therefore,
we have the $R$-module and $\overline{R}$-module isomorphisms
$$
\begin{array}{ccc}
 \begin{array}{cccc}
  \Psi_{H}: & R^\alpha  & \longrightarrow & R[H]  \\
    & (u_1,\ldots, u_{\alpha}) & \longmapsto &
     \sum\limits_{i=1}^{\alpha}u_ih_i,
\end{array}   & \text{ and } &  \begin{array}{cccc}
  \overline{\Psi}_{K}: & \overline{R}^{\,\beta}  & \longrightarrow & \overline{R}[K]  \\
    & (\overline{u}_{\alpha+1},\ldots, \overline{u}_{\alpha+\beta}) & \longmapsto &
     \sum\limits_{i=\alpha+1}^{\alpha+\beta}\overline{u}_ik_i,
\end{array}
\end{array}
$$ respectively. Thus, the map
$$
\begin{array}{cccc}
  \Psi_{(H,K)}: & R^{\,\alpha}\times\overline{R}^{\,\beta} &\longrightarrow & R[H]\times\overline{R}[K]  \\
    & (\textbf{u},\,\overline{\textbf{v}}) & \longmapsto &
    \left(\Psi_{H}(\textbf{u}),\,\overline{\Psi}_{K}(\overline{\textbf{v}})\right)
\end{array}
$$ is an $R$-module isomorphism, which leads to the following definition.

\begin{Definition}  An
$R\overline{R}$-linear code $C$ of block-length $(\alpha, \beta)$
is an \textit{$R\overline{R}$-linear group code with respect to $(H, K)$},
if $\Psi_{(H,K)}(C)$ is an ideal of the product ring $R[H]\times
\overline{R}[K]$.
\end{Definition}

\begin{Proposition}\label{sep} Let $C$ be a an $R\overline{R}$-linear group code with respect to
$(H,K)$. Then $C=C_1\times \overline{C}_2$ where $\Psi_H(C_1)$ is an
ideal of $R[H]$ and $\overline{\Psi}_K(\overline{C}_2)$ is an
ideal of $\overline{R}[K]$.
\end{Proposition}

\begin{proof} We have $\Psi_{(H,K)}(C)=\mathcal{I}\times\overline{\mathcal{J}}$, where $\mathcal{I}$ and
$\overline{\mathcal{J}}$ are ideals of $R[H]$ and
$\overline{R}[K]$, respectively, since the ideals of the product
ring $R[H]\times \overline{R}[K]$ are product of ideals of $R[H]$
and $\overline{R}[K]$. Thus, $C=C_1\times \overline{C}_2$, with
$\mathcal{I}=\Psi_H(C_1)$ and
$\overline{\mathcal{J}}=\overline{\Psi}_K(\overline{C}_2)$.
\end{proof}

\begin{Lemma}\label{equi}\cite[Theorem 3.9.]{GMS20} Let $\{C, D\}$ be an LCP of product group codes over a
finite chain ring. Then $C$ and $D^\perp$ are equivalent,
$D^\perp$ is the Euclidean dual of $D$.
\end{Lemma}
 Now we are in a condition to state a result that generalizes the result of Borello et al. \cite{BCW20} for our underlying space.
\begin{Theorem} Let $\{C,\,D\}$ be a pair of $R\overline{R}$-linear product group codes with respect to
$(H,K)$. Then $C$ and $D^\perp$ are equivalent.
\end{Theorem}

\begin{proof} From Proposition \ref{sep}, we have: $C=C_1\times \overline{C}_2$ and $D=D_1\times
\overline{D}_2$, where
\begin{itemize}
    \item  $\Psi_H(C_1)$ and $\Psi_H(D_1)$ are ideals of $R[H]$;
    \item  $\overline{\Psi}_K(\overline{C}_2)$ and $\overline{\Psi}_K(\overline{D}_2)$ are ideals of $\overline{R}[K]$.
\end{itemize}
Lemma \ref{sep-lcp} implies $\{C_1,\, D_1\}$ and
$\{\overline{C}_2,\,\overline{D}_2\}$ are LCP of product group
codes over finite chain rings $R$ and $\overline{R}$,
respectively. By Lemma \ref{equi}, $C_1$ and $D_1^\perp$ are
equivalent, and on the other hand $\overline{C}_2$ and
$\overline{D}_2^\perp$ are equivalent. Corollary \ref{sep-dual}
says $D$ is separable and
$D^\perp=D_1^\perp\times\overline{D}_2^\perp$. For that, it
follows that $C$ and $D^\perp$ are equivalent.
\end{proof}



\begin{thebibliography}{99}

\bibitem{AS15} I. Aydogdu, and I. Siap, \emph{On $\mathbb{Z}_{p^r}\mathbb{Z}_{p^s}$-additive codes},  Linear and Multilinear Algbra, \textbf{63}(10) (2015)  2089--2012.

\bibitem{BBDF20} N. Benbelkacem, J. Borges, S.T. Dougherty, and C. Fern\'{a}ndez-C\'{o}rdoba, \emph{On {$\Bbb Z_2\Bbb Z_4$}-additive complementary dual codes and related LCD codes}, Finite Fields Appl.,  \textbf{62} (2020) 101622.

\bibitem{BCW20} M. Borello, J. Cruz, and W. Willems, \emph{A note on linear complementary pairs of group codes}. Discrete. Math. \textbf{343}(8) (2020) 111905.

\bibitem{BFMBB20} S. Bhowmick, A. Fotue-Tabue, E. Mart\'inez-Moro, R. Bandi, and S. Bagchi, \emph{Do non-free LCD codes over finite commutative Frobenius rings exist?}, Des. Codes Cryptogr.,  \textbf{88}  (2020) 825--840.


\bibitem{BFRR10} J. Borges, C.  Fern\'andez-C\'ordoba, J. Pujol, J. Rif\'a, M. Villanueva, \emph{$\mathbb{Z}_2\mathbb{Z}_4$-linear codes: Generator matrices and duality}, Des. Codes Cryptogr., \textbf{54} (2010) 167--179.

\bibitem{BFKM23} M. Bajalan, A. Fotue Tabue, J. Kabore, and E. Mart\'inez-Moro: \emph{Galois LCD codes over mixed alphabets}, Finite Fields Their Appl., \textbf{85} (2023) 102125.

\bibitem{CG16} C. Carlet, and S. Guilley : \emph{Complementary dual codes for counter-measures to side-channel attacks},  J. Adv. Math. Commun., \textbf{10}(1) (2016) 131--150.

\bibitem{CGOOS18} C. Carlet , C. G\"uneri, F. \"Ozbudak, B.  \"Ozkaya, and  P. Sol\`e  : \emph{On linear complementary pairs of codes}. IEEE Trans. Inform. Theory, \textbf{64}(1) (2018) 6583--6588.


\bibitem{DS09} S.T. Dougherty, and H. Liu, \emph{Independence of vectors in codes over rings}. Des. Codes Cryptogr., \textbf{51} (2009) 55--68.

\bibitem{GMS20} C. G\"uneri, E. Mart\'inez-Moro, and S.Sayici: \emph{Linear complementary pair of group codes over finite chain rings}. Des. Codes Cryptogr.,  \textbf{88} (2020) 2397--2405.

\bibitem{LL15} X. Liu, and H. Liu: \emph{LCD codes over finite chain rings}. Finite Fields Their Appl.,   \textbf{34} (2015)  1--19.

\bibitem{LL23a} X. Liu, and H. Liu: \emph{Linear complementary pairs of codes over rings}. Des. Codes Cryptogr.,  \textbf{89} (2021) 2495--2509.

\bibitem{LL23b} X. Liu, and H. Liu: \emph{LCP of group codes over finite Frobenius rings}. Des. Codes Cryptogr., \textbf{91} (2023) 695--708.

\bibitem{NS00} G. Norton, and  A. Salagean: \emph{On the structure of linear and cyclic codes over finite chain rings}. Appl Algebra Eng Commun Comput.,  \textbf{10} (2000) 489--506.


\end{thebibliography}
\end{document}